\documentclass[preprint,superscriptaddress]{revtex4-1}

\usepackage[pdftex]{graphicx}
\usepackage{amsmath}
\usepackage[amssymb,Gray]{SIunits}
\usepackage{xfrac}

\newcommand{\bra}[1]{\langle #1 \hspace{-2pt} \mid}
\newcommand{\ket}[1]{\mid \hspace{-1pt} #1 \rangle}
\newcommand{\abs}[1]{\ensuremath{\left\vert #1 \right\vert}}
\newcommand{\D}{\mathrm{d}}

\begin{document}

% Use the \preprint command to place your local institutional report
% number in the upper righthand corner of the title page in preprint mode.
% Multiple \preprint commands are allowed.
% Use the 'preprintnumbers' class option to override journal defaults
% to display numbers if necessary
%\preprint{}

%Title of paper
\title{Optomechanical test of the Schr{\"o}dinger--Newton equation}

% repeat the \author .. \affiliation  etc. as needed
% \email, \thanks, \homepage, \altaffiliation all apply to the current
% author. Explanatory text should go in the []'s, actual e-mail
% address or url should go in the {}'s for \email and \homepage.
% Please use the appropriate macro foreach each type of information

% \affiliation command applies to all authors since the last
% \affiliation command. The \affiliation command should follow the
% other information
% \affiliation can be followed by \email, \homepage, \thanks as well.
%\author{}
%\email[]{}
%\homepage[]{Your web page}
%\thanks{}
%\altaffiliation{}
%\affiliation{}
%
\author{Andr\'e Gro{\ss}ardt}
\email[]{andre.grossardt@ts.infn.it}
\affiliation{Department of Physics, University of Trieste, 34151 Miramare-Trieste, Italy}
\affiliation{Istituto Nazionale di Fisica Nucleare, Sezione di Trieste, Via Valerio 2, 34127 Trieste, Italy}
\author{James Bateman}
\email[]{jbateman@soton.ac.uk}
\author{Hendrik Ulbricht}
\email[]{h.ulbricht@soton.ac.uk}
\affiliation{School of Physics and Astronomy, University of Southampton, SO17 1BJ, United Kingdom}
\author{Angelo Bassi}
\email[]{bassi@ts.infn.it}
\affiliation{Department of Physics, University of Trieste, 34151 Miramare-Trieste, Italy}
\affiliation{Istituto Nazionale di Fisica Nucleare, Sezione di Trieste, Via Valerio 2, 34127 Trieste, Italy}

%Collaboration name if desired (requires use of superscriptaddress
%option in \documentclass). \noaffiliation is required (may also be
%used with the \author command).
%\collaboration can be followed by \email, \homepage, \thanks as well.
%\collaboration{}
%\noaffiliation

\date{\today}

\begin{abstract}
The Schr{\"o}\-din\-ger--New\-ton equation has been proposed as an experimentally testable alternative to quantum
gravity, accessible at low energies. It contains self-gravitational terms, which slightly modify the quantum dynamics.
Here we show that it distorts the spectrum of a harmonic system. Based on this effect, we propose an optomechanical
experiment with a trapped microdisc to test the Schr{\"o}\-din\-ger--New\-ton equation, and we show that it
can be realized with existing technology.
\end{abstract}

% insert suggested PACS numbers in braces on next line
% insert suggested keywords - APS authors don't need to do this
%\keywords{}

%\maketitle must follow title, authors, abstract, \pacs, and \keywords
\maketitle

% If in two-column mode, this environment will change to single-column
% format so that long equations can be displayed. Use sparingly.
%\begin{widetext}
% put long equation here
%\end{widetext}

%%%%%%%%%%%%%%%%%%%%%%%%%%%%%%%%%%%%%%%%%%%%%%%%%%%%%%%%%%%%%%%%%%%%%%%%%%%%%%%%%%%%%%%%%%%
%%%  START  DOCUMENT  %%%%%%%%%%%%%%%%%%%%%%%%%%%%%%%%%%%%%%%%%%%%%%%%%%%%%%%%%%%%%%%%%%%%%
%%%%%%%%%%%%%%%%%%%%%%%%%%%%%%%%%%%%%%%%%%%%%%%%%%%%%%%%%%%%%%%%%%%%%%%%%%%%%%%%%%%%%%%%%%%

What is the gravitational field of a quantum system in a spatial superposition state?
The seemingly most obvious approach, the perturbative quantization of the gravitational field in analogy
to electromagnetism, makes it alluring to reply that the space-time of such a state must also be in a
superposition.
The non-renormalizability of said theory, however, has also inspired the hypothesis that a quantization of the 
gravitational field might not be necessary after all~\cite{Rosenfeld:1963,Mattingly:2005,Carlip:2008}.
Rosenfeld already expressed the thought that the question whether or not the gravitational field must be
quantized can only be answered by experiment:
\emph{``There is no denying that, considering the universality of the quantum of action, it is very tempting
to regard any classical theory as a limiting case to some quantal theory. In the absence of empirical evidence,
however, this temptation should be resisted. The case for quantizing gravitation, in particular, far from being
straightforward, appears very dubious on closer examination.''}~\cite{Rosenfeld:1963}

Adopting this point of view, an alternative approach to couple quantum matter to
a classical space-time is provided by a fundamentally semi-classical theory~\footnote{It needs to be stressed,
once again, that in this approach the coupling between gravity and quantum matter is \emph{fundamentally}
semi-classical. This should not be confused with models where Eq.~\eqref{eqn:sce} is considered an approximation
to some quantum theory of gravity~\cite{Anastopoulos:2014a}.}; that is by replacing the source term in Einstein's
field equations for the curvature of classical space-time, energy-momentum, by the \emph{expectation value} of the
corresponding quantum operator~\cite{Moller:1962,Rosenfeld:1963}:
\begin{equation}
\label{eqn:sce}
R_{\mu \nu} + \frac{1}{2} g_{\mu \nu} R = \frac{8 \pi G}{c^4} \,
\bra{\Psi} \hat{T}_{\mu \nu} \ket{\Psi} \,.
\end{equation}
Of course, such presumption is not without complications.
For instance, in conjunction with a no-collapse interpretation of quantum mechanics it would be in blatant contradiction
to everyday experience~\cite{Page:1981}. Moreover, the nonlinearity that the backreaction of quantum matter with
classical space-time unavoidably induces cannot straightforwardly be reconciled with quantum nonlocality in a
causality preserving manner~\cite{Eppley:1977,Gisin:1989}. Be that as it may, there is no consensus about the
conclusiveness of these arguments~\cite{Mattingly:2005,Kiefer:2007,Albers:2008}.
The enduring quest for a theory uniting the principles
of quantum mechanics and general relativity gives desirability to having access to hypotheses which could be put to an
experimental test in the near future.

In the nonrelativistic limit, the assumption of fundamentally semi-classical gravity yields a
nonlinear, nonlocal modification of the Schr{\"o}\-din\-ger  equation, commonly referred to as the
Schr{\"o}\-din\-ger--New\-ton equation~\cite{Giulini:2012,Giulini:2014,Bahrami:2014}.
After a suitable approximation~\cite{Giulini:2014}, for the center of mass of a complex quantum system
of mass $M$ in an external potential $V_\text{ext}$ it reads:
\begin{subequations}\label{eqn:sn}\begin{align}
\mathrm{i} \hbar \frac{\partial}{\partial t} \psi(t,\mathbf{r})
&= \left( \frac{\hbar^2}{2\,M}\nabla^2 + V_\text{ext} +  V_g[\psi] \right)\,\psi(t,\mathbf{r}) \\
V_g[\psi](t,\mathbf{r}) &= -G \, \int \D^3 r' \, \abs{\psi(t,\mathbf{r'})}^2 \, I_{\rho_c}(\mathbf{r}
- \mathbf{r'}) \,.\label{eqn:vg}
\end{align}\end{subequations}
The self-gravitational potential $V_g$ depends on the wavefunction, and hence renders the equation nonlinear.
The function $I_{\rho_c}$, which models the mass distribution of the considered system, will be defined below.

The Schr{\"o}\-din\-ger--New\-ton equation has primarily been discussed in the context
of gravitationally induced quantum state reduction~\cite{Diosi:1984,Penrose:1998}. Its relevance for a possible
experimental test of the necessity to quantize the gravitational field was pointed out by Carlip~\cite{Carlip:2008}.
First ideas how to test such kind of nonlinear, self-gravitational effects focused on the spreading of a free
wavefunction in matter-wave interferometry
experiments~\cite{Carlip:2008,Giulini:2011,Meter:2011,Giulini:2013,Colin:2014,Arndt:2014,Eibenberger:2013}.
Recently, an experimental test has been proposed by Yang et\,al.~\cite{Yang:2013}, based on the internal dynamics
of a squeezed coherent ground state of a micron-sized silicon particle in a harmonic potential.

\begin{figure}
\includegraphics[scale=.5]{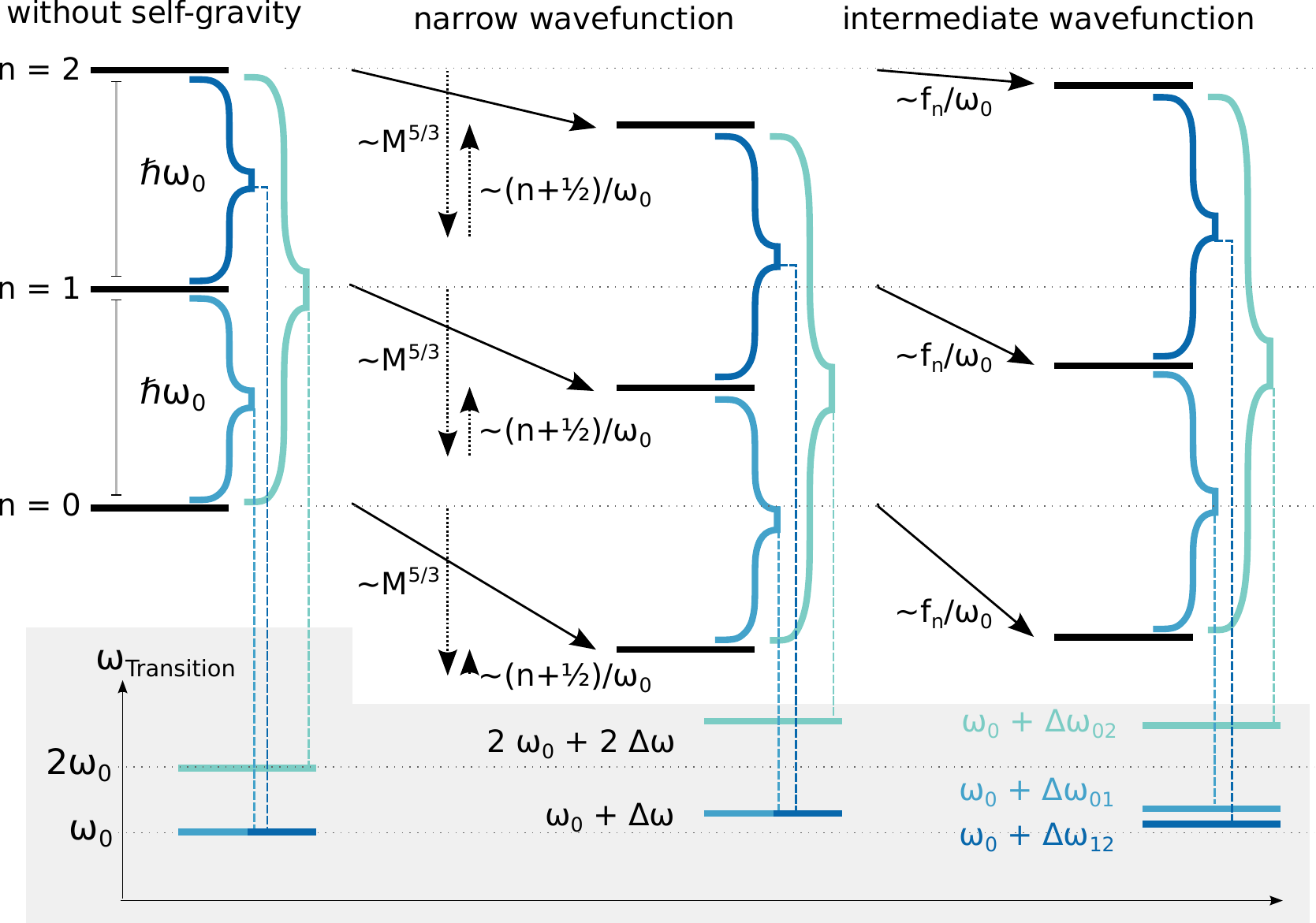} 
\caption{Schematic overview of the effect of the Schr{\"o}\-din\-ger--New\-ton equation on the spectrum.
The top part shows the first three energy eigenvalues and their shift due to the first order perturbative
expansion of the Schr{\"o}\-din\-ger--Newton potential. The bottom part shows the resulting spectrum of transition
frequencies. In the narrow wavefunction regime (middle part), all energy levels are shifted down by an $n$-independent
value minus an $n$-proportional contribution that scales with the inverse trap frequency. In the 
intermediate regime, where the wavefunction width becomes comparable to the localization length scale of the nuclei,
this $n$-proportionality does no longer hold, leading to a removal of the degeneracy in the
spectrum.\label{fig:overview}}
\end{figure}

Here we propose a novel experiment, studying the spectrum of such a harmonically trapped microparticle.
The advantage of our proposal is that it needs neither squeezing nor state tomography.
Therefore, despite the fact that effects are of the same order of magnitude than those studied in
Ref.~\cite{Yang:2013}, the prospects of its practical realization in the soon future are significantly better.

We show that for a suitable choice of mass and frequencies the energy levels become state dependent,
due to the self-gravity term $V_g[\psi]$ in Eq.~\eqref{eqn:sn}, thus resulting in a non-degenerate line spacing
of the energy levels of the harmonic oscillator. This is schematically depicted in Fig.~\ref{fig:overview}.

\section{Theory}
In matter-wave interferometry, even large molecules can approximately be considered as single, pointlike
particles, meaning that the wavefunction is wide in comparison to the extent of the considered quantum system.
In this case, the function $I_{\rho_c}$ in Eq.~\eqref{eqn:sn} reduces to the Coulomb-like potential
$M^2/\abs{\mathbf{r} - \mathbf{r'}}$. This changes if the wavefunction for the center of mass cannot be treated as wide,
as it is the usual situation in optomechanical experiments. In that instance, one must take the mass distribution
$\rho_c$ of the constituents relative to the center of mass into account. The general shape of $I_{\rho_c}$ is
\begin{equation}\label{eqn:i-rho}
I_{\rho_c}(\mathbf{d}) = \int \D^3 u \, \D^3 v \,
\frac{\rho_c(\mathbf{u}) \, \rho_c(\mathbf{v} - \mathbf{d})}{\abs{\mathbf{u} - \mathbf{v}}} \,.
\end{equation}
It has been pointed out by Yang et\,al.~\cite{Yang:2013} that, for a sufficiently narrow wavefunction, the crystalline
structure of matter becomes significant. Provided that the atomic mass density can be modeled by
a Gaussian distribution, and the microparticle as a whole has a spherical structure of radius $R$ much larger than
the extent of the wavefunction, we get approximately~\cite{Grossardt:2015b}
\begin{equation}\label{eqn:I-lattice-exp}
I_{\rho_c}(d) \approx \frac{6\,M^2}{5\,R} + \frac{M\,m}{d}\,\mathrm{erf}\left( \frac{d}{\sqrt{2}\,\sigma} \right)\,.
\end{equation}
Here, and in the following, we denote by $m$ the atomic mass, and by $M$ the mass of the whole microparticle.
The localization of the nuclei, $\sigma$, is related to the Debye--Waller B-factor by $\sigma = 2\pi\,\sqrt{B}$.
Values for $B$ at different temperatures can be found in Refs.~\cite{Sears:1991,Gao:1999} for most elemental crystals.

The external potential is now supposed to be a harmonic trap with frequency $\omega_0$ in $x$-direction, and we
assume the wavefunction to separate in the three spatial dimensions~\footnote{Note that according to the full 
Eq.~\eqref{eqn:sn} even an initially separable state will evolve into a nonseparable one. Only with the 
approximation~\eqref{eqn:energy-correction} the self-gravitational potential becomes separable.}.
The additional self-gravitational potential leads to
a shift of the energy levels from their unperturbed values at $E_n^{(0)} = \hbar\omega_0\,(1/2+n)$.
If the system is in a stationary state of the trap, and the mass is such that the self-gravitational potential is
weak, as is the case in any realistic experimental situation, the energy shift is well approximated by a first-order
perturbative expansion in the gravitational constant $G$:
\begin{equation}\label{eqn:energy-correction}
\Delta E_n = \bra{\psi^{(0)}_n} V_g[\psi^{(0)}_n](\mathbf{r}) \ket{\psi^{(0)}_n} + \mathcal{O}(G^2)\,.
\end{equation}
Strictly speaking, this is a twofold approximation, first by taking the unperturbed state $\psi^{(0)}_n$ as the source
of the gravitational potential, which renders $V_g$ a linear potential, and then applying ordinary perturbation theory.
Inserting the energy eigenstates of the harmonic oscillator, and introducing the dimensionless parameter
$\alpha = 2\,\sigma\,\sqrt{M\,\omega_0/\hbar}$, the energy shift can be written
\begin{equation}
\Delta E_n = -\frac{G\,\hbar\,m}{4\,\sigma^3\,\omega_0} \, f_n(\alpha)\,,
\end{equation}
with the state, mass, and frequency dependent functions $f_n$ yet to be determined.

We first consider the situation of Ref.~\cite{Yang:2013}, where the potential $V_g$ was simplified further by
taking the limit of a narrow wavefunction, and Taylor expanding $I_{\rho_c}$ to quadratic order in
$\abs{\mathbf{r}-\mathbf{r'}}$. In this case, the function $f_n$ takes
the form
\begin{equation}\label{eqn:energy-correction-narrow}
f_n^\text{narrow}(\alpha) = \;\propto M^{5/3} \omega_0
- \frac{4}{3}\,\sqrt{\frac{2}{\pi}}\,\left(n + \frac{1}{2} \right) \,.
\end{equation}
Obviously, the transition frequency $\omega_0$ between adjacent energy levels will not be affected by the
$n$-independent part, while the $n$-proportional term leads to a shift
\begin{equation}
\label{eqn:transition-frequency-shift-narrow}
\Delta \omega_\text{SN} = \sqrt{\frac{2}{\pi}}\,\frac{G \, m}{3\,\omega_0\,\sigma^3} \,. 
\end{equation}
This frequency shift is, however, independent of $n$, therefore leaving the degeneracy of the spectrum intact,
according to which all energy transitions with the same $\Delta n$ correspond to the same spectral line at
$\Delta n\, (\omega_0 + \Delta \omega_\text{SN})$.

The experimental situation of Ref.~\cite{Yang:2013}, where the frequency is high enough to allow for the
wavefunction to be approximated as narrow, hides the true behavior of the energy levels.
This becomes evident if we consider the scenario where the width of the wavefunction is comparable to the
localization $\sigma$ of the nuclei, and hence the quadratic approximation becomes inaccurate.
The wavefunction width is characterized by the factor $\alpha$ defined above, which relates the width of the
ground state to $\sigma$. Small values, $\alpha \ll 1$, correspond to wide wavefunctions and large values,
$\alpha \gg 1$, to narrow ones.

For now, consider only the case where the trap frequency is lowered in one dimension, but the wavefunction is
kept narrow in the remaining two.
Then, in the intermediate regime, where $\alpha$ is of the order of unity, the function $f_n$ can be approximated
(see Ref.~\cite{Grossardt:2015b} for a more thorough derivation)
\begin{widetext}
\begin{subequations}\begin{equation}
\label{eqn:fn-split-gauss}
f_n(\alpha) \approx \text{const.} + \alpha^3\,\sqrt{\frac{2}{\pi}} \, \int_0^\infty \D \zeta \,
\exp \left(-\frac{\alpha^2\,\zeta^2}{2}\right) \, P_n(\alpha\,\zeta) \,
\left(\frac{\mathrm{erf}\left(\sqrt{2}\,\zeta\right)}{2\,\zeta} - \sqrt{\frac{2}{\pi}}\right) \,,
\end{equation}
where ``const.'' refers to $n$-independent terms, and the polynomials $P_n$ are defined by
\begin{equation}
P_n(z) = \frac{\mathrm{e}^{-z^2/2}}{\sqrt{2\pi}\,(2^n \, n!)^2} \,
 \int_{-\infty}^\infty \D \xi \, \mathrm{e}^{-2\xi^2} \,H_n\left( \xi \right)^2
\left( \mathrm{e}^{2\,z\,\xi} \, H_n\left( \xi - z \right)^2
+ \mathrm{e}^{-2\,z\,\xi} \, H_n\left( \xi + z \right)^2 \right) \,,
\end{equation}\end{subequations}
\end{widetext}
with the Hermite polynomials $H_n$.
The integrals in $f_n$ can be evaluated analytically for low $n$~\footnote{We have been using \textsc{Mathematica}
to obtain analytical results for the functions $f_n$ up to $n=14$. The computation time increases
super-proportionally with $n$, making it infeasible to obtain results for higher $n$.}.
One ends up with the frequency shift
\begin{subequations}\label{eqn:delta-omega}\begin{align}
\Delta \omega^\text{interm.}_{n_1 n_2} &= \Delta \omega_\text{SN} \, g_{n_1 n_2}(\alpha) \\
g_{n_1 n_2}(\alpha) &= \frac{3}{8}\,\sqrt{2\,\pi}\,\left(f_{n_1}(\alpha)-f_{n_2}(\alpha)\right) \,,
\end{align}\end{subequations}
which now depends not only on the difference $\Delta n$ but explicitly on $n_1$ and $n_2$.
This is the effect we are interested in, which can be observed experimentally.
$\Delta \omega_\text{SN}$ contains the material properties, while $g_{n_1 n_2}(\alpha)$ depends only on
the total mass $M$ and trap frequency $\omega_0$.

\begin{figure}
\includegraphics[scale=.6]{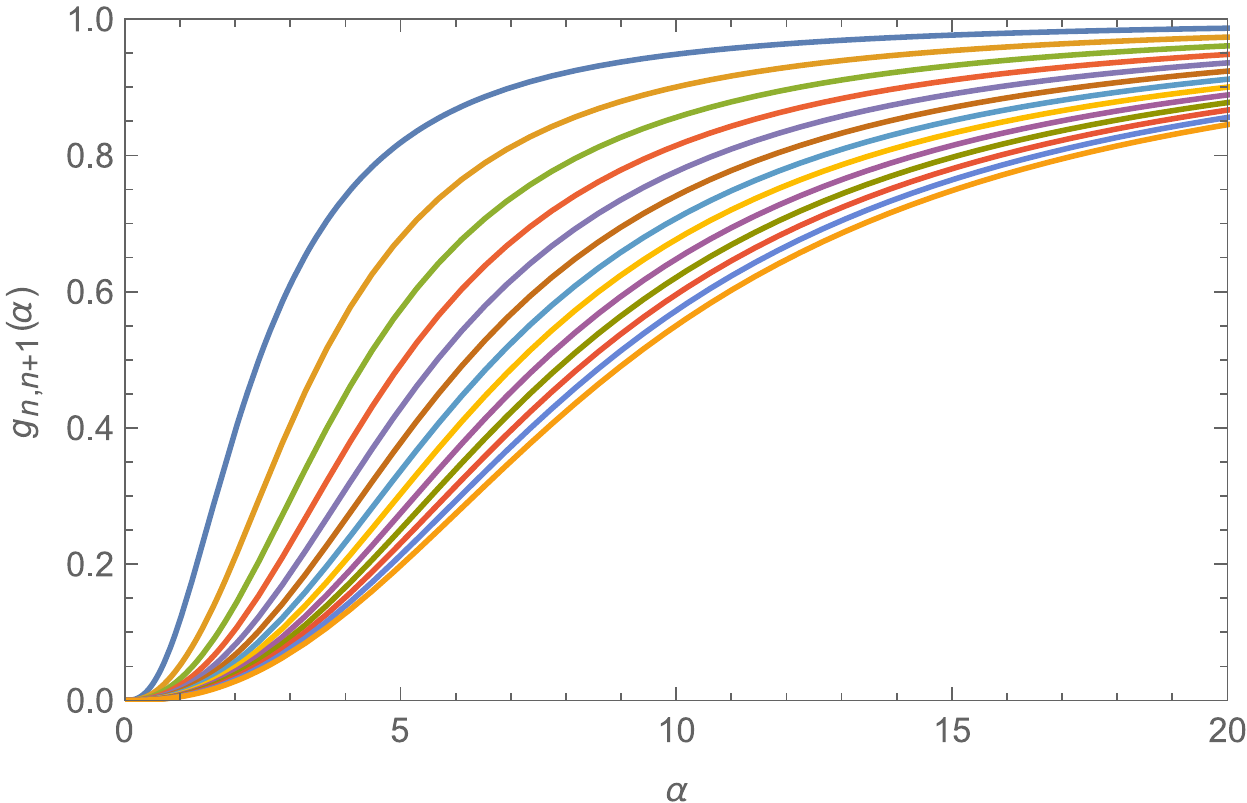}
\caption{Plot of the coefficient function $g_{n_1 n_2}(\alpha)$ for the lowest 13 transitions between harmonic
oscillator eigenstates with $\Delta n = 1$. $n$ increases from top to bottom, the blue curve belonging to $n_1=0$.
The parameter $\alpha$ characterizes the wavefunction width. Small values of $\alpha$ (wide wavefunctions) correspond
to smaller masses (for a given frequency) and therefore weaker self-gravity. In the limit of narrow wavefunctions
(large $\alpha$) the lines become degenerate again, all $g_{n_1 n_2}$ tending to $\Delta n = 1$.\label{fig:plot_gn}}
\end{figure}

The functions $g_{n_1 n_2}$ are plotted in Fig.~\ref{fig:plot_gn} for $\Delta n = 1$ and $n_1$ ranging from 0 to 12.
As one can see in the plot, while they tend to zero for small $\alpha$ (wide wavefunctions) and to $\Delta n$
for large $\alpha$ (narrow wavefunctions), for values $1 \lesssim \alpha \lesssim 10$ there is a significant dependence 
on the actual state, leading to a substantial deviation from the degenerate structure of the spectrum.

A fully three-dimensional analysis, giving up the approximation of a wavefunction that is narrow in two
dimensions, can only be obtained numerically. For an axially symmetric wavefunction that is in the ground state
for the transverse directions the details can be found in App. A of Ref.~\cite{Grossardt:2015b}. There it has been
shown that the effect stays the same in quality and order of magnitude also in this three-dimensional situation.

\section{Proposal for experiment} 
\begin{figure}
\includegraphics[width=0.45\textwidth]{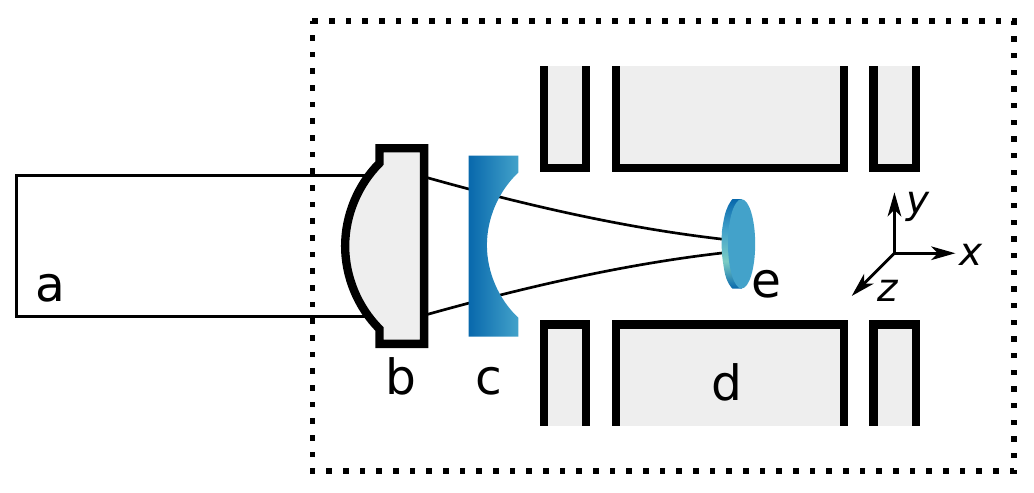}
\caption{Schematic of proposed experiment showing a) optical access, b) mode-matching lens, c) concave cavity mirror,
d) ion trap electrode structure, and e) electrically levitated superconducting disc. The disc acts as a concave cavity
mirror and forms an optomechanical cavity with mirror (c) to read the position of the disc. The apparatus is enclosed
in a dilution refrigerator to reduce thermal noises. The longitudinal direction, which will be used to investigate
the self-gravity modification of the mechanical harmonic oscillator energy levels, is along the
$x$-axis.\label{fig:schematic}}
\end{figure}

We propose to measure this effect by interrogating optomechanically the motion of a
single micron-sized superconducting osmium mirror in a dilution refrigeration-cooled linear Paul ion trap, as shown
schematically in Fig.~\ref{fig:schematic}. The Schr{\"o}dinger--Newton effect will be probed with the longitudinal
motion ($x$-direction in Fig.~\ref{fig:schematic}) of the trapped osmium microdisc.

While best known for atomic and molecular ions, the first demonstration of Paul traps included a $20\,\upmu{\rm m}$
particle~\cite{wuerker1959electrodynamic}, and a recently renewed interest has shown single micron-scale particles at
100\,Hz frequencies with single-charge resolution~\cite{gerlich2003molecular,schlemmer2004interaction}.
Longitudinal frequency can be orders of magnitude lower than the transverse~\cite{drewsen2000harmonic,landa2012modes},
which accurately embodies our theoretical description that the wave function remains narrow in two dimensions.

\subsection{Magnitude of the expected effect}
\begin{figure}
\includegraphics[scale=.6]{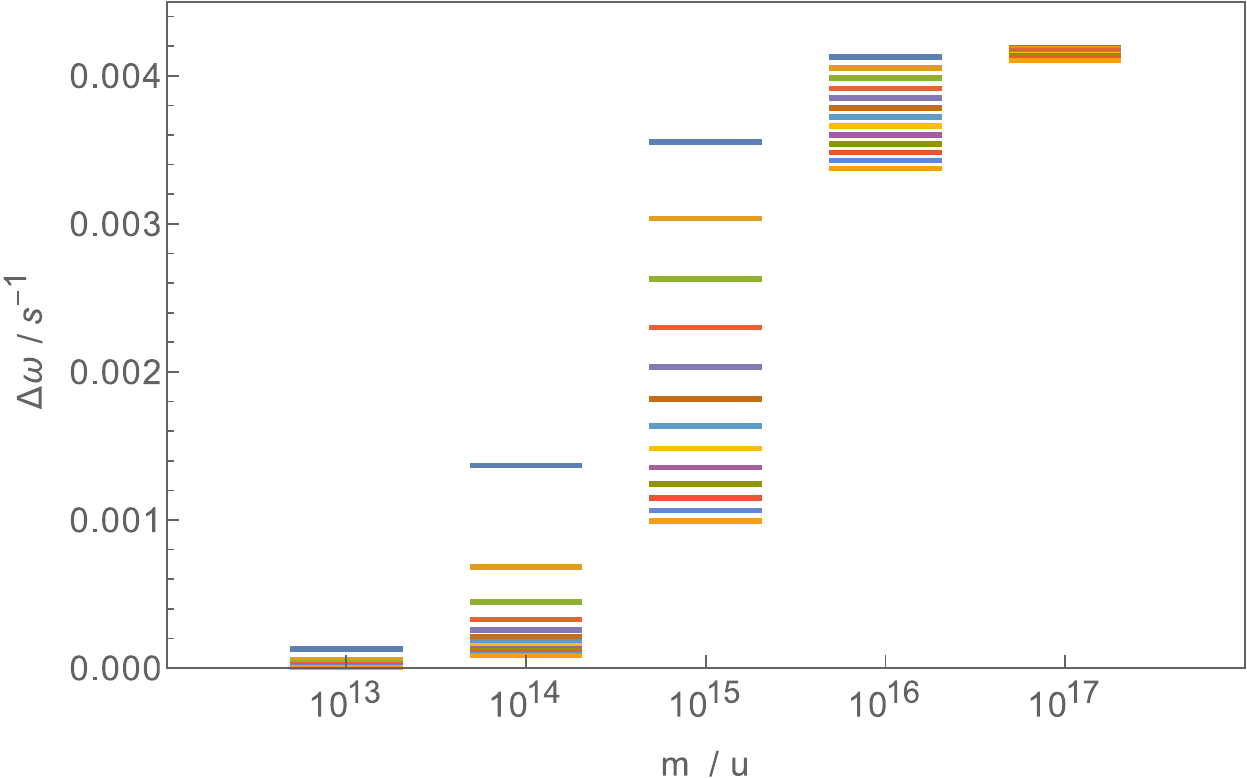}
\caption{The resulting frequency spectrum for osmium at $T = 100\,\milli\kelvin$
 at trap frequency $\omega_0 = 2\pi\times10\,\second^{-1}$ with $\Delta n = 1$.
At low masses, self-gravity becomes negligible. At high masses all spectral lines are degenerate,
shifted by $\Delta \omega_\text{SN}$. The intermediate regime, where a significant splitting appears,
spans about 3 orders of magnitude in mass.\label{fig:plot_spectrum}}
\end{figure}

Requiring that the ground-state wavefunction be in the intermediate regime, $\alpha \approx 5$, yields
$M=\hbar/(2\sigma/\alpha)^2/\omega_0\approx 10^{16}\,\atomicmass/\left(\omega_0/2\pi\,{\rm s}^{-1}\right)$.
Values of $\Delta \omega_\text{SN}$ for selected materials can be found in Tab.~\ref{tab:elements}.
We choose osmium, which has superconducting critical temperature $T_{\rm C}=700\,{\rm mK}$, favorable Debye--Waller
B-factor~\cite{Gao:1999}, and offers the smallest particle for a given mass.
\begingroup
\squeezetable
\begin{table}
\begin{ruledtabular}
\begin{tabular}{lrrrr}
Material & $m$ / \atomicmass & $\rho / \gram\per\centi\meter\cubed$ & $\sigma$ / \pico\meter
& $\Delta \omega_\text{SN}$ / $\second^{-1}$ \\
\hline
Silicon & 28.086\textsuperscript{a} & 2.329\textsuperscript{a} & 6.96\textsuperscript{b} & 0.00246 / ($
\omega_0 / \second^{-1}$) \\
Tungsten & 183.84\textsuperscript{a} & 19.30\textsuperscript{a} & 3.48\textsuperscript{b} & 0.128 / ($
\omega_0 / \second^{-1}$) \\
Osmium & 190.23\textsuperscript{a} & 22.57\textsuperscript{a} & 2.77\textsuperscript{c} & 0.264 / ($
\omega_0 / \second^{-1}$) \\
Gold & 196.97\textsuperscript{a} & 19.32\textsuperscript{a} & 4.66\textsuperscript{b} & 0.0574 / ($
\omega_0 / \second^{-1}$)
\end{tabular}
\end{ruledtabular}
\raggedright
\textsuperscript{a}Reference~\cite{particledatagroup},\quad
\textsuperscript{b}Reference~\cite{Sears:1991},\quad
\textsuperscript{c}Reference~\cite{Gao:1999}
\caption{Relevant material properties for selected elements. $\sigma = 2\pi\,\sqrt{B}$ is defined as in the text and 
depends on the Debye--Waller B-factor. Values are at  $T = 100\,\milli\kelvin$. We give $\Delta \omega_\text{SN}$,
which determines the magnitude of the effect according to Eq.~\eqref{eqn:delta-omega}, depending on
the trap frequency $\omega_0$.\label{tab:elements}}
\end{table}
\endgroup
The spectral lines are plotted for different mass osmium particles at a trap frequency
$\omega_0 = 2\pi\times10\,\second^{-1}$ in Fig.~\ref{fig:plot_spectrum}.
The split between adjacent spectral lines scales with $1/\omega_0$, just like the mass for which the effect is most 
pronounced.
For $M=10^{15}\,\atomicmass$, corresponding to an osmium particle (density $\rho=22.57 \,{\rm g}/{\rm cm}^3$) of
diameter $5.2\,\micro\meter$, we predict a frequency splitting  $\Delta f\sim 0.1\,\milli\hertz$.

A spherical particle would have radius comparable with typical laser wavelengths, making it impractical for use in a
concave-convex cavity. Instead, the superconductor should be a thin disk $\gtrsim 3\,\micro\meter$ in diameter and
$\sim 1\,\micro\meter$ thick~\footnote{While we considered a spherical particle in the theory section, the shape of
the particle contributes only to the $n$-independent shift of the energy levels. For the effect we are interested in, 
only the mass distribution of the nuclei inside the particle is relevant.}. Finesse in cavities with wavelength-scale 
mirrors is limited by mirror size and orientation stability~\cite{kleckner2010diffraction}.

\subsection{Competing effects}
All competing heating effects must produce a damping rate which is low compared with the frequency shift we expect:
sub-Hertz heating rates have been demonstrated in room-temperature conventional traps~\cite{poulsen2012efficient}
and cryogenics reduces heating rates substantially, as demonstrated in microfabricated 
traps~\cite{labaziewicz2008suppression};
rates depend strongly on microfabrication processes, and improvements to the levels we require are likely.
Indeed, already a cryogenically-cooled conventional trap should fulfill our requirements with existing technology.

Decoherence routes include interaction with blackbody radiation:
however, for radiation frequencies below the superconductor gap energy there is no 
absorption~\cite{dressel2013electrodynamics},
so by ensuring environmental temperature much less than superconductor critical temperature,
interaction with ambient photons is exclusively via Rayleigh scattering.
To reach this regime requires dilution refrigerator temperatures and we assume $T=100\,{\rm mK}$.
Using the particle as a mirror, not a sub-wavelength particle, means there is negligible Rayleigh scattering of laser
light~\cite{guccione2013scattering}, which would otherwise be a decoherence mechanism. Rayleigh scattering rate
decreases sharply for long wavelengths, and we find negligible probability of even a single scattering
event~\footnote{See supplemental material}.
Decoherence from collisions with background gas is also negligible for the low pressures
($P\lesssim 10^{-10}\,{\rm mbar}$) in the UHV cryo-pumped environment~\cite{romero2012quantum}.

\subsection{Optomechanical readout}
Lack of a long-lived internal state in our particle, such as those employed in sideband resolved manipulation of
trapped atomic ions, means that many techniques for engineering Fock states in ion traps~\cite{leibfried2003quantum}
cannot be applied here. Instead we appeal to optomechanics: inspired by cantilever
experiments~\cite{aspelmeyer2014cavity}, where a microfabricated cantilever provides one mirror of a high-finesse
cavity, we use the levitated metallic superconducting particle of sufficient reflectivity as one mirror in a tightly
focused, plano-concave, high optical Q cavity. Cantilever position can be monitored at the shot-noise 
limit~\cite{purdy2013observation},
and protocols have been proposed to prepare~\cite{Galland2014} and reconstruct~\cite{vanner2015towards} Fock states
in these systems. 

There are experiments with levitated hybrid nanoparticle ion-optomechanical systems, albeit with sub-wavelength 
particles~\cite{millen2014optomechanical}. Creation of Fock states of mechanical motion or phonon number states for 
$10^{14}\,\atomicmass$ particles has previously been proposed for a magnetically-levitated superconducting microsphere 
coupled to a quantum circuit~\cite{romero2012quantum}, while its experimental demonstration is yet to be shown.

\section{Conclusion}
We predict a new self-gravity effect to shift the energy states of a massive mechanical oscillator. We propose an
experimental scenario which makes use of the best parts of the mature technology of levitation of ions in Paul traps,
the cavity enhanced optical position readout and the use of superconducting materials to avoid competing heating effects
for instance by black body radiation. Our proposal is based on technology available today. However, the different
techniques have to be combined. It does not require the preparation and tomography of squeezed states of the mechanical
harmonic oscillator. Mechanical Fock states need to be prepared and read out, which is technology under intense
development as of today and has yet to be achieved. The self-gravity induced energy level splitting effect survives also
for states above the mechanical vacuum state, and hence particularly for thermal states, which means cooling to the
ground state is not required as such. Therefore our proposed experiment is feasible  with existing technology.

% If you have acknowledgments, this puts in the proper section head.
\begin{acknowledgments}
The authors gratefully acknowledge funding and support from the John Templeton foundation (grant 39530).
AG acknowledges funding from the German Research Foundation (DFG). JB and HU acknowledge support from the
UK funding body EPSRC (EP/J014664/1). HU acknowledges financial support from the Foundational Questions
Institute (FQXi). AB acknowledges financial support from the EU project NANOQUESTFIT, INFN, and
the University of Trieste (grant FRA 2013).
\end{acknowledgments}

%

%%%%%%%%%%%%%%%%%%%%%%%%%%%%%%%%%%%%%%%%%%%%%%%%%%%%%%%%%%%%%%%%%%%%%%%%%%%%%%%%%%%%%%%%%%%
%%%  SUPPLEMENTAL  MATERIAL  %%%%%%%%%%%%%%%%%%%%%%%%%%%%%%%%%%%%%%%%%%%%%%%%%%%%%%%%%%%%%
%%%%%%%%%%%%%%%%%%%%%%%%%%%%%%%%%%%%%%%%%%%%%%%%%%%%%%%%%%%%%%%%%%%%%%%%%%%%%%%%%%%%%%%%%%%

\onecolumngrid
\clearpage
\begin{center}
\textbf{\large Supplemental Material for
``Gravitational fine structure of harmonically trapped particles according to the
Schr{\"o}\-din\-ger--New\-ton equation''}
\end{center}
%%%%%%%%%% Merge with supplemental materials %%%%%%%%%%
%%%%%%%%%% Prefix a "S" to all equations, figures, tables and reset the counter %%%%%%%%%%
\setcounter{equation}{0}
\setcounter{figure}{0}
\setcounter{table}{0}
\setcounter{section}{0}
\setcounter{page}{1}
\makeatletter
\renewcommand{\thesection}{S\arabic{section}}
\renewcommand{\theequation}{S\arabic{equation}}
\renewcommand{\thefigure}{S\arabic{figure}}
\renewcommand{\bibnumfmt}[1]{[S#1]}
\renewcommand{\citenumfont}[1]{S#1}
%%%%%%%%%% Prefix a "S" to all equations, figures, tables and reset the counter %%%%%%%%%%

%\begin{bibunit}[apsrev4-1]

\section*{Rayleigh scattering of blackbody radiation by superconducting microdisc}\label{suppsec:rayleighscatter}

For radiation frequencies below the superconductor gap energy~\cite{dressel2013electrodynamics-supp} there is no
absorption; by ensuring the radiation temperature is much less than the gap energy, we can assume that no photons with
energy exceeding the gap are encountered.
Under these conditions, the polarizability of the sub-wavelength particle is $\chi=3V$ where $V$ is the volume of the
particle.
The Rayleigh scattering cross-section is $\sigma_R=k^4\chi^2/6\pi$ where $k=\omega/c$ is the wavenumber.
Integrating over the blackbody energy density $u(T,\omega)$ we find the total rate of photon scattering
\[
\Gamma_R=\int_0^\infty\D\omega \frac{\sigma_R c u(T,\omega)}{\hbar\omega} = 30\,720\pi^5\zeta(7)\times
c\,\frac{\chi^2}{\lambda_{\rm T}^7}\approx 10^7\times c\, \frac{\chi^2}{\lambda_{\rm T}^7}
\]
where $\zeta$ is the Riemann zeta function ($\zeta(7)\approx 1.01$) and $\lambda_{\rm T}=hc/k_BT$ is the typical
wavelength of the thermal radiation at temperature $T$.  For $T=100\milli{\rm K}$, we find $\lambda_{\rm T}\approx 14\centi\metre$.
Therefore, for both a sphere of $V=\frac{4}{3}\pi (1\micro\metre)^3$
and a disc of $V=\frac{\pi}{4}\,(1\micro\metre)\,(3\micro\metre)^2$ the scattering rate is
$\Gamma_R\sim 10^{-12}\second^{-1}$.

%%%
% Bibliography for supplemental material
%%%
% !! requires additional run of: bibtex bu1.aux
%%%
\def\bibsection{\section*{References to supplemental material}}
\let\endnote=\ignore
%

% avoid second output of bibliography (dirty hack)
\makeatletter
\let\thebibliography=\ignore
\let\endthebibliography=\endignore
\makeatother

\end{document}